\documentclass{cernrep}
\usepackage[numbers]{natbib}
\usepackage{natbib}

\begin{document}
\title{Think Physics, Think Man: Barriers to Women's Participation in Physics Education (OPINION)}
\author{Eliot Walton}
\institute{Monash University}

\maketitle

\section{Framework}

This text focuses on women because the research that is available tends to fall into binary gender categories. This work is not generalisable to other genders without further research. Furthermore, while the discussion draws on my experience as a trans-woman, I have omitted a discussion of trans-specific issues (such as transition leave and transphobia) for generality. In my estimation, the most common situation for a female physics student is that she is cis-gender and one of a small number among a class primarily composed of cis-men. I seek to explore two possible barriers to her learning, the notion of competency and gender dynamics in groups. 

\section{Competency}

Having transitioned from male-presenting to female-presenting over the course of second and third year, I can attest to the degree to which being perceived as female in physics is synonymous with being perceived as incompetent. My particular situation was extremely bizarre, where the more I studied (hence the more I knew) the less I was assumed to know by male peers. Rather than focusing on mansplaning \cite{men_explain} which is a particular type of condescending explanation given by a man to a woman where an unreasonable degree of incompetence is ascribed to the woman, I will focus on why women are assumed to be incompetent in physics and how the way that physics is taught is likely disadvantage women.
\\
\\
Educational experiments also link a student's belief in themselves about their ability to learn material with success in learning the material \cite{GOOD2003645, best-model}. So self-belief has a tangible effect on actual competence as well as perceived confidence. Therefore, an individual's idea of what skills and types of intelligence are necessary for a discipline have an effect on their success in the discipline.
\\
\\
Leslie \textit{et al.,} argue that these perceptions can help account for the distribution of women across academic disciplines \cite{doi:10.1126/science.1261375, doi:10.1126/science.aaa9632, doi:10.1126/science.aaa9892}. In particular, they argue that raw intellectual talent, which is associated more with men than women \cite{doi:10.1126/science.1261375}, helps to explain the lack of women in physics, mathematics and philosophy since these disciplines are assumed to require such skills. If one must be 'innately' talented to succeed, and women are taught that innate talent tends not to reside in them, then it is not surprising that women feel alienated from the field. It is likely that this alienation encourages women to experience "imposter syndrome", which is most commonly understood to be a feeling of "intellectual fraudulence" \cite{fraud}, accompanied by the conviction that success is unearned and future endeavours are likely to fail \cite{2022}. A 2022 study highlights that minorities (such as women) in 'brilliance' oriented disciplines experience higher levels of imposterism, which has negative consequences on their future career \cite{2022}.
\\
\\
These results must be carefully interpreted. This is an additional barrier which women face compared to men, but it does not negate all other barriers (such as the lack of role models and being a numerical minority in classes). However, this is an effect of which teachers should be mindful when presenting class material; emphasis on individual brilliance (regardless of whether it is by men or by women) produces a set of expectations about the requirements for success which alienate women from the field. 
\\
\\
The other care that must be taken is to avoid adopting the how-to-ask-for-a-promotion solution where the problem is treated as being with the women rather than with the way that physics is taught. It is very easy to conclude that women need to be be more like men in order to be successful in physics and that we should teach them accordingly. However, this is incorrect because it places the burden of addressing the inequality on women, and argues they must be less like themselves to succeed. Rather, we should meet women where they are and acknowledge that, by the time a student is at university, she has already internalised such beliefs. To un-learn them is a slow process which requires that we re-think the way we teach physics.

\section{Group Dynamics}

Gender dynamics operate in all interpersonal relationships. If they are unnoticed by some that is likely because they have become familiar and have therefore receded from view but they are often obvious to individuals whose gender is conspicuous. Since men are the universal default \cite{Invisible_women} what is unmarked is assumed to be male. Men make up the majority of people seen or heard in films, on television or over radio \cite{Invisible_women}. They make up the majority of world leaders, CEOs and professors \cite{women_and_leadership}. Therefore man-to-man dynamics are the most commonly represented gender dynamics and this conspires with the male default to make the interactions between men appear as the standard rather than one possible combination of genders. For example, we have the Australian Football League (AFL), and the Australian Football League Women (AFLW); women are an afterthought if they are a thought at all. The dynamics in a physics classroom between members of a group are often male, focusing on competition and individualism.
\\
\\
This is exclusive of women, but that is not the main point of concern. The main point of concern is how these dynamics interact with women when women are introduced \cite{truth_about_women_talking}. Barbara and Gene Eakings found that, in a recording of seven university faculty meetings, that men spoke more than women, with one exception and that the longest comment by a woman was shorter than the shortest comment by a man \cite{Y}. Analysing online discussions about linguistics between men and women revealed that men's messages tended to be twice as long as women's messages \cite{Susan}.
\\
\\
While I know of no data which gives these numbers for physics classrooms, it seems reasonable to hypothesise that a similar pattern will hold. While speaking is not the only form of engagement with class material, in interactive and flipped classrooms it is a major one. Flipped classrooms are those where traditional lectures and homework are reversed. Speaking is the verbal equivalent of taking up space, and women are already taught to minimise and be conscious of the space they occupy. It is therefore expected that these formats disadvantage women. Furthermore, while men measurably speak more in these studies, women are stereotyped as speaking more \cite{women_talking_more} and this effects the perception of who speaks more. 
\\
\\
My father recently remarked to me about the number of women economists consulted on the news and pointed out that I would be annoyed if they were all men economists. The implication being that these situations are equally inequitable. Despite the perceived abundance women, when measured by the Global Media Monitoring Project, which has evaluated the world's media for the representation of women every five years since 1995, their 2021 report found that women only make up 25\% of persons heard, read about or seen in radio, newspapers or television \cite{GMMP_2020}. This represents a single point improvement since 2015 and the first improvement since 2010. Despite this incremental progress, when Caroline Criado-Perez campaigned in 2013 for a female historical figure on the back of English banknotes many men were upset with one man writing "but women are everywhere now!" \cite{Invisible_women}.
\\
\\
While wrong about the numbers, this man's experience is enlightening because he (and those who think like him) experience even a minor intrusion of women as an inequity. This is important because it highlights that men's experience of how many women there are, how often women speak and for how long is biased against women. Therefore, even there are numerically fewer women and they measurably speak less, men may still experience this as them 'being everywhere'. This conclusion is empirically supported by research from the Global Institute for Women's Leadership (GIWL), which found that medical professionals tended to over-estimate women's true representation in several different areas of medicine and in different roles \cite{medical}.
\\
\\
The GIWL research indicates that a man's degree of overestimation predicted lower support for gender equality initiatives (with a p-value less than 0.001). Women's (over)estimations were unrelated to their level of support (with a p value of 0.92) \cite{medical}. This study establishes an empirical relationship between medical practitioners' perception of women's representation and their willingness to support gender equality initiative but makes no attempt to explain why such a relationship exists. It is reasonable to assume that such men see these initiatives as inequitable or unnecessary and that this is a consequence of their biased perception against women. 
\\
\\
While these data have not been collected for physics it is reasonable to hypothesise that they would show a similar pattern. A recent study of 
 27 self-identified progressive white-male physics faculty and graduate students found that these men despite their self-image as equity champions acted in ways that maintained the unequal \textit{status quo} \cite{https://doi.org/10.48550/arxiv.2210.03522}. If this is the case for progressive men, then it is reasonable to assume those who do not think of themselves as such will be even more inhibitive to women's progress and initiatives. In physics, the degree to which women are perceived as intruding may be greater than other fields because of the degree of male-domination. An example of this thinking was provided by Prof. Alessandro Strumia in 2018 where he claimed that men were discriminated against in the hiring practices of Center for European Nuclear Research (CERN) \cite{Alessandro}. These claims were made despite women representing 21\% of CERN staff in 2019 \cite{Cern_hiring}. 
\\
\\
He subsequently published a heavily criticised paper attempting to demonstrate this perceived inequality. This paper did "not provide a convincing understanding of the literature or the methods [of the field of gender studies]" \cite{science_article} according to Cassidy Sugimots, a researcher on gender disparities in science \cite{Larivière_Ni_Gingras_Cronin_Sugimoto_2013}.

\section{The Glass Cliff}

When all other solutions fail, try women. An educational class in honours electrodynamics operated via this principle; when a male colleague and I failed to solve a problem we turned to the quiet observer, the other woman in the class, and asked directly for her input. In transitioning I have found particular patterns of thinking which originate in my socialisation including many of the patterns of thinking critiqued here. 
\\
\\
This honours class shows an example of the glass-cliff phenomenon which is the tendency to appoint women to precarious leadership positions and was first identified in 2004 \cite{women_and_leadership}. It is reasonable to think this contributed to the 2022 appointment of Elizabeth Truss as Prime Minister of the United Kingdom, and may also have affected the election of Giorgia Meloni in Italy since this effect has been measured repeatably in women's leadership \cite{women_and_leadership}. Perhaps this effect appears also in physics classrooms, where more space and credence tends to be made for and lent to women's input when problems are more difficult.
\\
\\
While this seems to directly contradict the notion that women are less competent there is no contradiction. Women are sought out of desperation rather than deference. Discussions of the glass cliff especially in the case of male-dominated environments often include a discussion of conscious or unconscious malice; that the woman is set-up to fail or expected to fail. I do not suggest that men in physics ask their female colleagues in bad faith (although this is possible). However, it is important to be wary of the possibility of bias confirmation; if a woman is unable to do the harder problems when she is turned to this may depress her confidence and reinforce her peers' beliefs (conscious or otherwise) about women's inability to do physics.
\\
\\
Collaboration is an important aspect of research and teaching attempts to reflect this in its assessment. While it is reasonably well understood that group projects will have uneven workloads between members and that this is reflective of real world practices the way that this type of assessment effect women verses men should be considered when setting such tasks. In workplace teams, women tend to do a disproportionate amount of the labour \cite{women_groups} and in some fields, such as economics, they tend to be given less credit for their collaborative work \cite{254946}. It is not clear if these results are applicable to a physics classroom; however, I suspect that the amount of work women do will be disproportionate in group tasks. This suspicion comes from discussions with undergraduates and is an extension of glass cliff thinking. A student may not be incorporated if the work is trivial, and she may be given more when it is difficult solely due to glass-cliff thinking. Combining this with the disproportionate load women carry in workplace teams, I expect that, when her contributions are integrated, she will do the majority of the work, or she will do none of it.

\section{Conclusion and Further Avenues}

\subsection{Conclusion}

A woman in a physics classroom faces obvious and non-obvious barriers to her learning. The story of physics research is told in a way that emphasises the innate ability of the individual which biases physics teaching and success towards men. It is likely that this has a measurable affect on women's mastery of the  class material. Gender dynamics affect physics classrooms and it is expected that these  dynamics disadvantage women compared to men especially in the case of flipped class rooms. However, even though women speak less and are fewer, they still may be perceived as too numerous leading to a reduction in men's support of gender equality, which may correlate with a reduction in their awareness of the ways in which their behaviour excludes women. When women are included in groups they are treated differently to men by men, who either expect them to do all of the work or none of it, generally undervalue their contributions and only turn to them as an emergency resource. Therefore, it is expected that women will have differential outcomes to men when passing through a physics curriculum.

\subsection{Further Avenues}

This paper does not explore:
\begin{itemize}
\item Intersectionality
\item Differential behaviour and expectations
\item Gendered violence 
\item The de-emphasis of women in science history
\item Teaching methodology and the built environment
\item Role models
\item Mentoring programs
\item Health (especially access to feminine hygiene products)
\item The effect of flexible learning (or lack thereof)
\item The effect and distribution of unpaid domestic work
\end{itemize}

\noindent all of which are worthy of exploration. I acknowledge that many expectations here are based on inference and speculation and that further research is required to support or refute many claims before action may be taken on them. This document's primary purpose is educational and discursive.  

\section*{Acknowledgements}

I respectfully acknowledge the Boonwurrung people of the Kulin nation as the traditional owners and custodians of the land on which I live and work. I pay my respects to Elders past, present and emerging and support the establishment of an Aboriginal and Torres Strait Islander Voice. I wish to thank Monash's Women and Non-Binary People in Physics and Astronomy (WiNPA) for their ongoing support and companionship. I also wish to thank the Monash Equity, Diversity and Inclusion (EDI) committee for their support. In particular, I wish to thank Rowina Nathan, Maddy Howell, Javira Altmann, Sabrina Rock and Dr Mark Walton for their insightful discussion, comments and support. I also wish to thank Prof. Ulrik Egede for his willingness to support this work and its publication.

\newpage

\bibliographystyle{unsrtnat}
\bibliography{main}

\end{document}